\def\year{2022}\relax
\documentclass[letterpaper]{article} 
\usepackage{aaai22}  
\usepackage{times}  
\usepackage{helvet}  
\usepackage{courier}  
\usepackage[hyphens]{url}  
\usepackage{graphicx} 
\usepackage{natbib}  
\usepackage{caption} 

\usepackage{booktabs}

\DeclareCaptionStyle{ruled}{labelfont=normalfont,labelsep=colon,strut=off} 
\usepackage{algorithm}
\usepackage{algorithmic}

\usepackage{newfloat}
\usepackage{listings}
\floatstyle{ruled}
\newfloat{listing}{tb}{lst}{}
\floatname{listing}{Listing}
\title{Understanding Toxicity Triggers on Reddit in the Context of Singapore}
\author{
    Yun Yu Chong, Haewoon Kwak
}
\affiliations{
    Singapore Management University, Singapore\\

    yunyu.chong.2018@scis.smu.edu.sg, haewoon@acm.org
}

\begin{document}

\maketitle

\begin{abstract}
While the contagious nature of online toxicity sparked increasing interest in its early detection and prevention, most of the literature focuses on the Western world. In this work, we demonstrate that 1) it is possible to detect toxicity triggers in an Asian online community, and 2) toxicity triggers can be strikingly different between Western and Eastern contexts.
\end{abstract}

\section{Introduction}

With the widespread use of the internet and social media, online interactions have become a vital part of social lives today.
From social media to online games, people actively interact with each other through online channels, seeking social support and a sense of community~\cite{oh2014does}.  
With the global COVID-19 pandemic, people have spent more time online. 

These online interactions, however, sometimes may become harmful~\cite{mondal2017measurement,10.1145/2702123.2702529}. 
``Toxic'' comments, which are often rude, vulgar, or discriminative, are  prevalent on social media. According to~\cite{duggan2017online}, 41\% of US citizens experienced online harassment as of 2017.
Moreover, these toxic comments are contagious~\cite{kim2021distorting}; exposure to toxic comments leads to subsequent toxicity. 
This contagious nature of online toxicity sparked increasing interest in the communities developing methods for early detection and prevention of hate speech~\cite{karan2019preemptive,paschalides2020mandola,an2021predicting}. 
One such effort is detecting \emph{toxicity triggers}, which are the origins that initiate toxic discussions~\cite{almerekhi2019detecting,alshamrani2020investigating,almerekhi2020these}. Despite this growing interest, most studies have focused on the Western world, mainly due to the availability of natural language processing tools and data. 

In this work, we demonstrate that 1) it is possible to detect toxicity triggers in an Asian online community, and 2) toxicity triggers can be strikingly different between Western and Eastern contexts by comparing the triggers in the Western and Asian online communities.  
For the Asian community, we focus on Singapore, a multicultural country where English is the dominant language. For the Western community, we study New York City, a multicultural city with a comparable population. 
Specifically, we pose the following research questions: 

\begin{itemize}
    \item RQ1: What are the characteristics of toxic comments in online discussions in the context of Singapore? 
    \item RQ2: What are the common toxicity triggers, i.e., starting points that lead to toxic conversations, in online discussions in the context of Singapore? 
    \item RQ3: How are those toxicity triggers similar or different between Singapore and New York City?
\end{itemize}

\section{Related Work}
Since the 1990s, researchers have been proposing methods for detecting toxic comments. From Microsoft’s rule-based hate speech comment detector ``Smokey''~\cite{spertus1997smokey} to traditional classifiers like Naïve Bayes applied to features from Bag-of-Words (BoW)~\cite{davidson2017automated}, word embeddings, and more advanced Natural Language Processing (NLP) methods~\cite{djuric2015hate}, there have been many efforts in this area. The introduction of pre-trained models and transformers like BERT for text classification~\cite{zhao2021comparative} have further proved the success of toxic comment detection.

Almerekhi et al.~\shortcite{almerekhi2019detecting,almerekhi2020these} proposed a novel approach to understanding how toxic conversations are initiated. Toxicity triggers, which they defined as ``comments that incur direct toxic responses,'' were detected by training a Long Short-Term Memory (LSTM) neural network model using a combination of features. Their approach focuses on the 10 Reddit communities with the most subscribers as of 2017, which yielded diverse but limited results. Alshamrani et al.~\shortcite{alshamrani2020investigating} examined 14,500 videos and their 7.3 million 
comments from 30 popular YouTube news channels by using an ensemble of classifiers. They found an association between specific topics (e.g., religion and crime-related news) and toxicity of comments. 


\section{Data Collection}

Reddit is a popular social media platform with 52M daily active users as of October 2020~\cite{patel2020reddit}. 
On Reddit, the r/singapore~\cite{rsingapore} subreddit focuses on Singaporean experiences and news. With 473K members as of January 2022, it is the biggest online community for Singaporeans on the web.
There are other online communities that focus on Singapore (e.g., HardwareZone.sg Forum), but they are all significantly smaller than r/singapore. 
For this study, we collect all posts and comments on r/singapore between January 2021 and July 2021 \cite{Baumgartner_Zannettou_Keegan_Squire_Blackburn_2020}. 
We perform data cleaning to remove irrelevant comments that contain no meaningful text, such as image-only or links-only comments, deleted comments, and replies from  moderators or bots. After cleaning, we have 643,913  comments.

\section{Toxic Comments}

\subsection{Toxic Comment Detection}

The first step to addressing RQs is to identify the toxic comments. 
While it is challenging to train a custom model from scratch, a good range of well-performing pre-trained models is available for consideration due to the high interest in online toxicity research. 
We find 20 text classification transformers for detecting English toxic comments from  Hugging Face's model repository.
After filtering out some poor-performing models by testing a manually curated small set of comments, we feed 2,000 randomly sampled test cases obtained from Jigsaw’s toxic comment classification challenge~\shortcite{kaggle2018jigsaw} into the remaining eight models. 
Given that Hugging Face models return the toxicity score from 0 to 1, we define a comment as “toxic” when a score $>=$ 0.8 and  “non-toxic” when a score $<=$ 0.2. 
Only three out of eight models achieve accuracies of higher than 0.9, with F1-scores higher than 0.65, signaling that they can be good baselines. 




To ensure that these models perform well in the Singaporean context, we conduct a manual evaluation using 200 randomly sampled comments from r/singapore from May 2021, the month with the highest number of comments (146K).
These comments are labeled  by three annotators currently living in Singapore and compared against the model predictions.
Although the accuracies of all three models are between 0.7 and 0.8, the F1-scores are below 0.3.

We then test Perspective API~\cite{perspectiveapi}, which is widely used for toxicity research~\cite{fortuna2020toxic}. It shows an accuracy score of 0.94 and an F1-score of 0.6, which is better than Hugging Face models. As a result, we decide to use Perspective API to estimate the toxicity of all 643,913 comments in our dataset.


\begin{table}[h]
\centering \small
\begin{tabular}{ccc}
\toprule
Category & Predicted Score & N of  Comments \\
\midrule
Toxic & $\geq$ 0.8 & 20,599 (3.2\%) \\
Non-Toxic & $\leq$ 0.2 & 477,124 (74.1\%) \\
Ambiguous & $>$ 0.2 and $<$ 0.8 & 134,724 (20.9\%) \\
Others & NA & 11,466 (1.8\%) \\
\bottomrule
\end{tabular}
\caption{Distribution of Toxicity scores}\label{tab:toxicity_score}
\end{table}

Perspective API successfully estimates toxicity for more than 98\% of the comments in the dataset. 
Similar to $\S$4.1, we categorize comments into “toxic” ($>=$ 0.8) or “non-toxic$<=$ 0.2) based on their score. The remaining comments are labeled as “ambiguous.” 
We find that 74.1\%, 20.9\%, and 3.2\% of the comments are non-toxic, ambiguous, and toxic, respectively (Table 1). 


We note some limitations in this step. First, because the API predicts the probability that a comment is toxic, we define a strict cut-off -- 0.8 and 0.2 for ``toxic'' and ``non-toxic,'' respectively -- while comments that fall outside the cut-off are considered to be ``ambiguous.''
Although some ambiguous comments are indeed toxic or non-toxic, eliminating them is necessary to avoid inducing noise into the data. 
A second limitation is that the API can only predict the toxicity of one supporting language at a time. 
Some of the comments in the ``Others'' (1.8\%) category are written in  more than one language, and so may contain non-English characters or words. 
As Singapore is a multi-ethnic society with four official languages, it would be a natural step to include this type of comment. We leave this for future work.

\subsection{Characterization of Toxic Comments}

To answer RQ1, we focus on the words that distinguish toxic comments from non-toxic ones.
We use a scaled F-Score~\cite{kessler2017scattertext}, which is the harmonic mean of magnitude-adjusted and standardized category-specific precision and category-specific term frequency. 
The Scaled F-Score ranges between -1 and 1, where a positive score indicates that the term is more associated with the target category, and a negative score indicates the reverse. 

\begin{figure}[h!]
\centering
  \includegraphics[width=80mm]{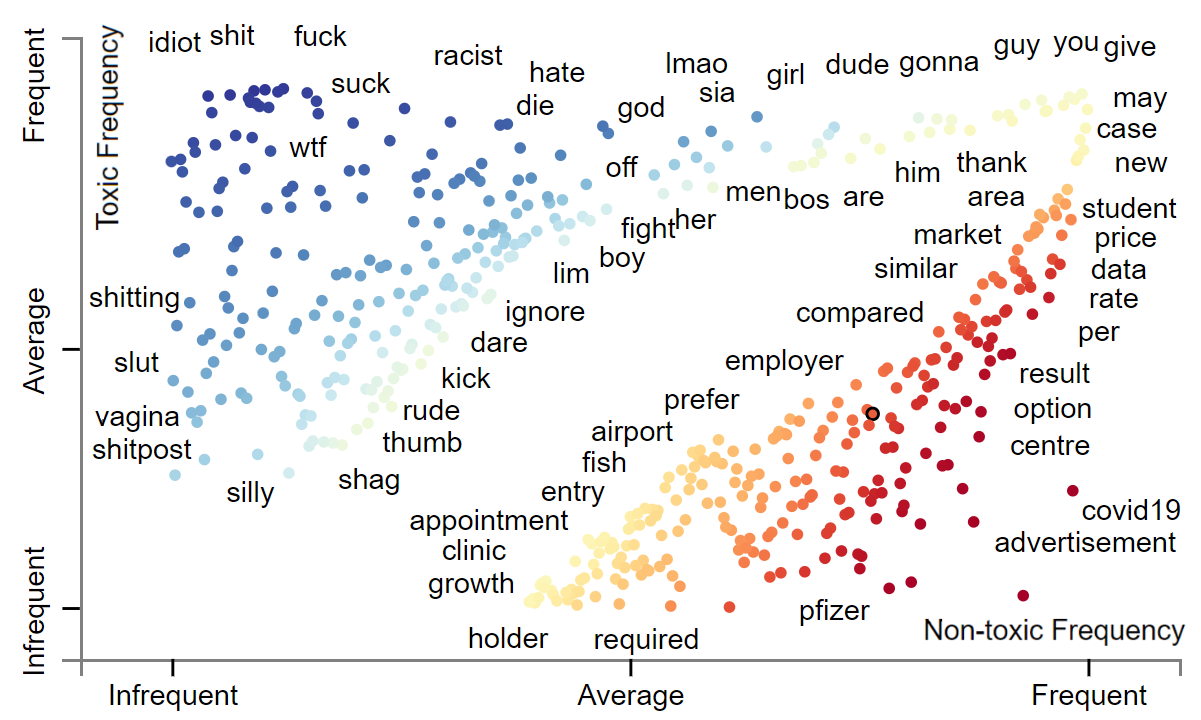}
  \caption{Frequent words in toxic and non-toxic comments}
  \label{fig:scattertext}
\end{figure}



Figure~\ref{fig:scattertext} shows the distinguishing words in toxic and non-toxic comments. 
The top words in toxic comments, as seen in the top left corner of Figure \ref{fig:scattertext}, are all swearwords, while the top words in non-toxic comments, as seen in the bottom right corner, are mostly about COVID-19. 
Also, we can see several themes that provide a glimpse of the possible toxicity triggers. 
Words about culture, race, and religion are prominent, and the three most dominant races in Singapore -- Chinese, Indian, and Malay -- are explicitly presented. 
The term “xenophobic” also further highlights the long-running issue of foreigners in Singapore. 
Multiple terms refer to the Singaporean government, such as “gov,” “govt,” “government,”  “gahmen,” and many acronyms of government-related persons or agencies, like “oyk” which stands for the Minister of Health, Mr. Ong Ye Kung. 
These terms are expected due to the strong focus on local news in r/singapore.



\section{Toxicity Triggers}


A toxicity trigger is defined as a non-toxic comment that has at least $n$ toxic child comments, meaning that it initiates subsequent toxic discussions. 
Following \cite{almerekhi2020these}, we first find toxicity triggers and non-triggers by checking their child comments. We then use them to fine-tune a pre-trained language model, Bidirectional Encoder Representations from Transformers (BERT), for predicting toxicity triggers. 
In order to try to filter out a random toxic comment unrelated to the initiator, we use $n=2$ instead of 1.
We further use all the 1,135 identified triggers ($n=2$) and the same number of randomly selected non-triggers to fine-tune BERT, using an 80/20 split. 
The prediction model of toxicity triggers achieves an F1-score of 0.79.

\subsection{Toxicity Trigger Analysis}

To address RQ2, we adopt a post-hoc approach to extract meaningful information from the learned model for toxicity prediction in the previous section. 
We use Shapley Additive exPlanations (SHAP)~\cite{NIPS2017_7062} to interpret the model. 
The main idea of SHAP is to find the average marginal contributions of each feature based on all permutations of the observations provided. 
We can identify the most triggering words in r/singapore based on the Shapley values of each word.
The top 20 words with the highest Shapley values are: creation, chad, uefa, noted, celebrations, disclosure, primary, philippines, influence, recommendations, again, registered, robbie, opium, ad, outstanding, centre, paralysis, squad, and male.

\begin{figure}[h!]
\centering
  \includegraphics[width=80mm]{sg_v2.pdf}
  \caption{Common words in comments containing top 500 triggers}
  \label{fig:top500}
\end{figure}

As these are word-level triggers, we need to consider the context to understand them. Figure~\ref{fig:top500} shows the common words in comments containing the top 500 triggers, which is considered to be the context of triggers. From observation, COVID-19 related topics are prominent, as seen from the words “Covid” and “case” in the word cloud, as well as topics related to schools, as identified from the words “student,” “school,” and “NTU” (Nanyang Technological University).

We perform a qualitative analysis of the comments that contain the top 500 trigger words.
As a result, we identify eight primary categories of toxicity triggers in r/singapore.

\begin{itemize}
    \item COVID-19 related news, regulations, and experiences: These are discussions related to COVID-19, including government policies such as social distancing measures, vaccinations, the reopening of borders, and updates about local infections. The toxic discussions relate to opinions or emotions about how the government is managing the pandemic. 
\item Racism and xenophobia: These triggers involve individuals or organizations that show racism and xenophobic behaviors, either on Reddit or offline. 
Following toxic comments target the entities involved.
\item Meritocracy and elitism in the local education system: These discussions mainly center on the design and structure of Singaporean education, which are related to racial issues, especially when the Special Assistance Plan  for bilingual students who were inculcated with traditional Chinese values~\cite{sap2016} or ``Chinese privilege''~\cite{velayutham2017races} is involved in the discussion. Following toxic comments express the unhappiness or disappointment with the education system in Singapore.
\item Mental health and stress: These comments are related to mental health, which is increasingly important in Singapore. According to \cite{mentalhealth_moh}, 13\% of the population reported experiencing depression or anxiety symptoms since the pandemic started. 
The triggers often contain news about local situations, government policies, and school support. Following toxic comments relate to negative personal experiences of seeking help and insufficient or inadequate mental health assistance.
\item Organization screw-ups: These triggers contain experiences or news about poor customer service or management of local businesses or organizations. Following toxic comments target the organizations involved, where most users are expressing their anger or disappointment.
\item Outrageous but trivial acts: These triggers are news about petty crimes or acts that are not illegal but violate social norms, such as misusing office supplies, catching seafood from the beach, and hoarding groceries. Following toxic comments target the people involved in these acts, showing disgust, embarrassment, or anger.
\item Advertisements gone wrong: These triggers are related to product or service advertisements, including government publicity materials, that did not serve their intended purposes. They often contain an image of the actual advertisement. For example, some triggers relate  to improper slogans used for an energy drink, unthoughtful images used for a supplement brand, and inappropriate use of memes for public announcements on social media. Following toxic comments commonly involve mockery or sarcasm toward the advertisements.
\item Exposés of scams: These triggers are related to personal experiences of being tricked by scammers. Following toxic comments are toward the scammers.
\end{itemize}

Two kinds of people are most frequently involved in the triggers: (1) local political figures and policymakers, and (2) local social media influencers. The triggers are often about what they claimed or did. The behaviors are sometimes trivial, such as riding a bicycle on an expressway, or significant, such as introducing new COVID-19 safety measures.

\section{Comparison of Toxicity Triggers in Singapore and New York City}

\subsection{Toxicity Triggers in New York City}

To further understand toxicity triggers in Singapore and how unique they are, we conduct a comparative study.
New York City (NYC) is chosen as a point of comparison to represent the Western point of view. The subreddit r/nyc (573k subscribers as of Jan 2022) is dedicated to discussing general NYC topics. 
The proportion of toxic comments per month (2.61\%) is comparable to r/singapore, making the comparison more direct. 

Similar to building a prediction model in r/singapore, the toxicity triggers with at least two toxic child comments in r/nyc are used to fine-tune another BERT model. The trigger prediction model for r/nyc obtains an accuracy of 0.76 and an F1-score of 0.77, which is comparable to r/singapore.

\subsection{Comparison of Toxicity Triggers}

Based on the Shapley values calculated for each word, the top 20 triggers from r/nyc are: cdc, photo, consent, snapped, apologized, john, flee, marijuana, tells, elimination, jeff, activists, who, chrysler, opens, proposes, twitter, lifted, stolen, and editorial. 
These triggers are quite different from those in r/singapore. 
We qualitatively analyze comments that contain the top 500 trigger words and identify five main topics, as we did for r/singapore. They are: 
\begin{itemize}
\item COVID-19 related news, regulations and experiences
    \item Protests or strikes against big companies
    \item  Elections and campaigns
\item Memes about politics or multinational corporation 
\item Investments
\end{itemize}



In r/nyc, the triggers are more about rights, power, and money, such as protests and elections, while in r/singapore, the focus is more on social or public issues, like school and mental health. News about local politicians and COVID-19 are prominent in both communities. However, the names of individual politicians seem to be more common in r/nyc, such as “cuomo,” “eric adams,” and “andrew yang,” while this is less common in r/singapore. 
It can also be observed that more focus is placed on what an individual politician, such as Governor Cuomo, said or did in the NYC community, while in Singapore, more focus is placed on the governments’ actions collectively. 
In r/nyc, most government-related issues discussed are politics and economics-related bills, while in r/singapore, they are about pandemic management.

The top triggers in r/singapore and r/nyc, based on the Shapley values, show stark differences. Among the top 200 triggers from each subreddit, we found that only 5\% of them are the same words. Even if we look at the top 1,000 triggers from each, only 30\% are overlapped. It demonstrates  unique triggers in Singapore (and New York City) and calls for follow-up  studies of more diverse countries to understand toxicity and its triggers in each corresponding context.


\section{Discussion and Conclusion}

This study has a strong contextual focus, where the online Singaporean community is studied using computational tools. Machine learning models and their interpretations deepen understanding of controversial issues in Singapore, which helps to learn what Singaporeans uniquely value.

Admittedly, there are some limitations in this work. As with all machine learning-based studies, the results observed are specific to the data used in the study. 
In particular, we collected the data for seven months during a period when local COVID-19 cases in Singapore were relatively stable, but issues related to COVID-19 were still prominent. 
Thus, the findings, such as government-related toxicity triggers, may or may not be representative of other periods, especially in normal times when the world is not in a pandemic. 
For future work, data from pre-pandemic times will be analyzed to better understand toxicity triggers in Singapore. 
Furthermore, this study is based on a general definition of online toxicity. 
There can be several nuances within the concept of toxicity, such as identity attacks that are hateful remarks towards a person’s identity such as race or gender, sexual harassment or references that are inappropriate comments involving body parts or sexual activities, misinformation that are comments which contain the fabrication of facts, amongst others. 
The triggers for specific toxicity can vary, and it would be worthwhile to compare them to gain deeper insights on online toxicity. 
Lastly, interpreting the context of the word-based triggers is particularly important, as our qualitative analysis shows that word-level triggers by themselves are not straightforward or sufficient to gain a complete picture. Additional topic modeling and causal effect estimation may provide novel and valuable insights.

All in all, our work focuses on understanding toxicity triggers in online discussions in the context of Singapore. Using Perspective API, we found that 1 in 33 r/singapore comments are toxic. By fine-tuning BERT, we built a custom language model to predict the probability of whether a comment is a toxicity trigger.
Our results showed that toxic comments often do not target other Redditors but the entities mentioned in the triggering comments. 
By comparing the triggers in r/singapore and r/nyc, we demonstrated   the uniqueness of the Singaporean community.
The more prominent topics observed in the discussions involve social issues that affect the public, such as education and mental health.

With the increasing digital connectivity worldwide, social media interactions would remain a vital part of human communication and connection. 
Research on toxicity triggers can help contain or mitigate the impacts of online toxicity, reducing the harm caused, and creating a more positive and enjoyable digital environment for all.

\section*{Acknowledgments}
Kwak gratefully acknowledges the support by D.S. Lee Foundation Fellowship awarded by Singapore Management University.

\bibliography{main_1075}


\end{document}